\documentclass[12pt,preprint]{aastex}

\usepackage{natbib}
\usepackage{amsmath}
\usepackage{units}

\DeclareMathOperator\erf{erf}
\DeclareMathOperator\F{F}
\DeclareMathOperator\U{U}

\shorttitle{Sparse On/Off Data: a Bayesian Analysis}
\shortauthors{Knoetig}

\begin{document}


\title{Signal discovery, limits, and uncertainties with sparse On/Off
measurements: an objective Bayesian analysis}


\author{Max L. Knoetig}
\affil{Institute for Particle Physics, ETH Zurich, 8093 Zurich,
Switzerland}
\email{mknoetig@phys.ethz.ch}







\begin{abstract}
For decades researchers have studied the On/Off counting problem, 
where a measured rate consists of two parts. One due to a signal 
process and another due to a background process, of which both 
magnitudes are unknown. While most frequentist methods are adequate 
for large count numbers, they cannot be applied to sparse data. Here 
I want to present a new objective Bayesian solution that only depends on three 
parameters: the number of events in the signal region, the number of 
events in the background region, and the ratio of the exposure for 
both regions. First, the probability of the hypothesis that the 
counts are due to background only is derived analytically. Second, 
the marginalized posterior for the signal parameter is also derived 
analytically. With this two-step approach it is easy to calculate 
the signal's significance, strength, uncertainty, or upper limit in 
a unified way. The approach is valid 
without restrictions for any count number including zero 
and may be widely applied in particle physics, cosmic-ray physics and high-energy astrophysics.
In order to demonstrate its performance I apply the method to gamma-ray burst data.
\end{abstract}


\keywords{gamma-rays: general --- methods: statistical}



\section{\label{sec:intro}Introduction}

Typical counting experiments measure a discrete set of events, such 
as the decay time of a particle. Such data are often \citep
{li1983analysis,cousins2008evaluation} modeled with the 
Poisson distribution given by
\begin{equation}
 P_{\text{P}}\left(N|\lambda\right)=\frac{e^{-\lambda}  \lambda^{N}}{N!} .
 \label{eqn:poisson}
\end{equation} 
This distribution connects the probability to observe N events to a 
nonnegative number of expected events $\lambda$, derived, for 
example, from a rate in a fixed time interval or a luminosity and a 
cross section. The Poisson distribution may be approximated by a 
Gaussian distribution when measuring many events. But when the data 
are sparse such an approximation is not good enough. Indeed, in some 
areas this is often the rule rather than the exception, such as 
high-energy astrophysics \citep{loredo1992promise}. In this paper a 
full Bayesian analysis of On/Off data, valid for any count number,
is presented.

\section{\label{sec:onoff}The On/Off Measurement}

In the On/Off problem one would like to infer a signal rate in the 
presence of an imprecisely known background. The measurement 
consists of the observation of $N_{\text{off}}$ events in a region 
chosen a priori to be signal free and the observation of 
$N_{\text{on}}$ events in a region of a potential signal in addition 
to the background.      

The notation comes from astronomy, where telescopes point 
\textit{on} and \textit{off} potential source regions. In particle 
physics, the \textit{off} region is taken in some region close to 
the signal region in the measured parameter \citep[typically called 
\textit{sideband}, e.g.][]{cousins2008evaluation} or without 
radioactive signal source near the detector.

The ratio of the exposure for both regions $\alpha$ is
assumed to be known with negligible 
uncertainty. In gamma-ray astronomy $\alpha$ is in the most simple cases
\begin{equation}
 \alpha=\frac{A_{\text{on}}  t_{\text{on}}}{A_{\text{off}}  t_{\text{off}}} ,
 \label{eqn:alpha}
\end{equation}
where $A$ stands for the size and $t$ for the exposure time of the 
regions. \citet{Berge2007} illustrate how to generalize 
Equation \ref{eqn:alpha} for complex acceptances.
Given $N_{\text{on}}$, $N_{\text{off}}$, and $\alpha$ the 
problem is to calculate the evidence for a signal and the posterior
distribution of the signal parameter.

Frequentist analyses, based on likelihood ratios and other methods, 
are widespread in particle physics \citep{cousins2008evaluation}
and in high-energy astrophysics \citep[as promoted by][]{li1983analysis}.
However, they often assume normal distributed random numbers and therefore
lose their foundation when applying them to low count numbers. 

\citet{gillessen2005significance} have proposed a 
Bayesian solution to the question whether the signal parameter is 
larger than zero. But they do not account 
for the alternative, simpler hypotheses asserting that all $N_{\text{on}}$
come from background only, i.e. no source. Therefore the result is an overestimation,
as pointed out by \citet{gregory2005bayesian}. 
There exists a Bayesian solution to 
compute the odds ratio of the two hypotheses \citep
{gregory2005bayesian}, but it adds an arbitrary parameter to the 
problem, depending on the prior (in particular its upper boundary), 
which makes the probability statement hard to interpret. 

\section{\label{sec:ana}Analysis}
In this paper an objective Bayesian analysis is presented. This is 
accomplished by using improper priors as a tool 
for producing proper posteriors representing our lack of knowledge.
Certainly, subjective Bayesian methods can have benefits in particular 
when a prior opinion is strongly held. 
One can gain sensitivity by using informative priors
that specify prior knowledge precisely, such as a prior 
source detection or a known background. 
But objective Bayesian methods should be used 
when one is interested in ``what the data have to say'' \citep{Irony1997}. 

The analysis presented here follows the recipe outlined by \citet
{caldwell2006signal} and may be considered an analytical special 
case.  \citet
{Agostini13} applied the recipe in order to analyze and set
stringent upper limits on 
the neutrinoless double beta decay of $^{76}$Ge. 
\citet{Kashyap2010} recently presented a similar frequentist method.
The analysis is performed in two steps. 
First, the probability that the observed counts are due to 
background only is calculated. If this is lower than a previously 
defined consensus value, the signal is said to be detected. Second, 
the signal contribution is estimated or an upper limit for the 
signal is calculated, depending on whether the detection limit has 
been reached.

\subsection{\label{sec:hyptest}Hypothesis Test}
Let $H_{\text{0}}$ denote the null hypothesis that \textit{the observed 
counts are due to background only}. The alternative hypothesis 
$H_{\text{1}}$ is that \textit{a signal process contributes to the counts}.
$H_{\text{1}}$ could be a bad model too, in case of systematic
uncertainties. It must be noted that sometimes this is the case when e.g. 
signal counts leak into the \textit{off} region. Nevertheless, in the following it is assumed that the 
systematic uncertainties are negligible and the two model set
of exclusive rival hypothesis 
$\{H_{\text{0}},H_{\text{1}}\}$ is complete.
By using Bayes' theorem one may calculate the conditional 
probability of
$H_{\text{0}}$ as
\begin{eqnarray}
 P\left(H_{\text{0}}|N_{\text{on}},N_{\text{off}}\right) &=&
\frac{P\left(N_{\text{on}},N_{\text{off}}|H_{\text{0}}\right) 
P_{\text{0}}\left(H_{\text{0}}\right)}{P\left(N_{\text{on}},N_{
\text{off}}\right)} , \label{eqn:1}
\end{eqnarray}
where $P\left(N_{\text{on}},N_{\text{off}}|H_{i}\right)$ is the 
conditional probability to observe the data, given the hypothesis 
$H_i$ and $P_{\text{0}}\left(H_{i}\right)$ is the prior probability 
for $H_i$.
For a set of exclusive rival hypothesis such that $\sum_i{P(H_{i})}=1$
and $P(H_{i} \land H_{j}) = 0$ for $i \neq j$, the law of total probability
gives
\begin{eqnarray}
 & & P\left(N_{\text{on}},N_{\text{off}}\right) = \sum_{i} P\left(N_{\text{on}},N_{\text{off}}|H_i\right) P_{\text{0}}\left(H_i\right) .
 \label{eqn:evidence1}
\end{eqnarray}
Furthermore, in continuously parametrized models the continuous counterpart 
of the law of total probability,
with sums replaced by integrals, gives
\begin{eqnarray}
 & & P\left(N_{\text{on}},N_{\text{off}}\right) = \nonumber \\ 
 & & \sum_{i}\int P\left(N_{\text{on}},N_{\text{off}}|\vec{
 \lambda}_i,H_i\right) P_{\text{0}}\left(\vec{\lambda}_i|H_i\right)d\vec{
 \lambda}_i P_{\text{0}}\left(H_i\right).
 \label{eqn:evidence}
\end{eqnarray}
The sum is made over the full set of hypotheses $H_i$ and the 
integration with respect to their parameters $\vec{\lambda}_i$.
By assuming the two hypothesis set $\{H_{\text{0}},H_{\text{1}}\}$ one can write 
Eqn. \ref{eqn:evidence} in terms of the expected number of signal events 
$\lambda_{\text{s}}$ and the expected number of background events 
$\lambda_{\text{bg}}$:
\begin{eqnarray}
 & & P\left(N_{\text{on}},N_{\text{off}}\right) =  \\ & & \int P
 \left(N_{\text{on}},N_{\text{off}}|\lambda_{\text{bg}},H_{\text
 {0}}\right) P_{\text{0}}\left(\lambda_{\text{bg}}|H_{\text{0}}
 \right)d\lambda_{\text{bg}} P_{\text{0}}\left(H_{\text{0}}
 \right) \nonumber \\ & & + \int P\left(N_{\text{on}},N_{\text
 {off}}|\lambda_{\text{s}},\lambda_{\text{bg}},H_{\text{1}}
 \right) P_{\text{0}}\left(\lambda_{\text{s}},\lambda_{\text
 {bg}}|H_{\text{1}}\right)d\lambda_{\text{s}}d\lambda_{\text
 {bg}} \nonumber \\ & & \times P_{\text{0}}\left(H_{\text{1}}
 \right). \nonumber
\end{eqnarray}
Here, 
$P\left(N_{\text{on}},N_{\text{off}}|\lambda_{\text{bg}},H_{\text{0}}
\right)$ and 
$P\left(N_{\text{on}},N_{\text{off}}|\lambda_{\text{s}},\lambda_{\text
{bg}},H_{\text{1}}\right)$ denote the conditional probabilities to 
measure the data.

Assuming that the number of signal events (if any) and the number of 
background events are independent Poisson-distributed random 
variables with means $\lambda_{\text{s}}$ and $\lambda_{\text{bg}}$, 
the expected number of events in the \textit{off} region is
\begin{equation}
 E\left(N_{\text{off}}\right)=\lambda_{\text{bg}} .
 \label{eqn:mean1}
\end{equation}
The expected number of events $E\left(N\right)$ in the \textit{on} region is, 
assuming the null hypothesis $H_{\text{0}}$ 
\begin{equation}
 E\left(N_{\text{on}}\right)= \alpha \lambda_{\text{bg}} ,
\end{equation}
or assuming $H_{\text{1}}$ 
\begin{equation}
 E\left(N_{\text{on}}\right)=\lambda_{\text{s}}+\alpha \lambda_{\text{bg}} .
 \label{eqn:mean2}
 \end{equation}
This yields for the conditional probabilities to measure the data or
\textit{likelihoods}
\begin{eqnarray}
& & P\left(N_{\text{on}},N_{\text{off}}|\lambda_{\text{bg}},H_{
\text{0}}\right) = \nonumber \\ & & P_{\text{P}}\left(N_{\text
{on}}|\alpha\lambda_{\text{bg}}\right) P_{\text{P}}\left(N_{
\text{off}}|\lambda_{\text{bg}}\right) ,
 \label{eqn:likelihood-1}
\end{eqnarray}
and
\begin{eqnarray}
& & P\left(N_{\text{on}},N_{\text{off}}|\lambda_{\text{s}},
\lambda_{\text{bg}},H_{\text{1}}\right) = \nonumber \\ & & P_{
\text{P}}\left(N_{\text{on}}|\lambda_{\text{s}}+\alpha\lambda_{
\text{bg}}\right) P_{\text{P}}\left(N_{\text{off}}|\lambda_{
\text{bg}}\right) .
\label{eqn:likelihood}
 \end{eqnarray}
The priors $P_{\text{0}}\left(\lambda_{\text{bg}}|H_{\text{0}}\right)$ and
$P_{\text{0}}\left(\lambda_{\text{s}},\lambda_{\text{bg}}|H_{\text{1}}\right)$
are chosen according to 
\textit{Jeffreys's rule} \citep[see][]{jeffreys1961theory,pdg} 
\begin{eqnarray}
P_{\text{0}}\left(\vec{\lambda}_i|H_{i}\right) & \propto & \sqrt{
\det\left[I\left(\vec{\lambda}_i|H_{i}\right)\right]} ,\\ I_{kl}
\left(\vec{\lambda}_i|H_{i}\right) & = & -E\left[\frac{\partial^
{2}\ln\, L\left(N_{\text{on}},N_{\text{off}}|\vec{\lambda}_i,H_{i}
\right)}{\partial\lambda_{k} \partial\lambda_{l}}\right] ,
\end{eqnarray}
where $I_{kl}$ denotes the \textit{Fisher information matrix}, $L$ 
the likelihood function (either Eqn. \ref
{eqn:likelihood-1} or \ref{eqn:likelihood}), $E$ the expectation 
value with respect to the model with index $i$ and $\vec{\lambda}_i$ 
its parameter vector. In Appendix \ref{app:jeffrey} I show that
\begin{eqnarray}
 P_{\text{0}}\left(\lambda_{\text{bg}}|H_{\text{0}}\right) &\propto& 
 \sqrt{\frac{1+\alpha}{\lambda_{\text{bg}}}} , \\ P_{\text{0}}
 \left(\lambda_{\text{s}},\lambda_{\text{bg}}|H_{\text{1}}
 \right) &\propto& \sqrt{\frac{1}{\lambda_{\text{bg}}
 \left(\alpha\lambda_{\text{bg}}+\lambda_{\text{s}}\right)}}.
\label{eqn:prior}
\end{eqnarray}
The On/Off Jeffreys's priors are improper, i.e.
integrate to infinity over the parameter space.
There is a debate among statisticians, concerning the use of improper priors 
in Bayesian model selection \citep{Berger2001}, as
the priors are only specified up to the proportionality constants $c_0, c_1$
which do not cancel out.
The probability of $H_{\text{0}}$ to be true given the measured 
counts is therefore
\begin{eqnarray}
P\left(H_{\text{0}}|N_{\text{on}},N_{\text{off}}\right) &=&  \frac{c_0 \gamma'}{c_0 \gamma'+ c_1 \delta'} \\
&=& \frac{ \gamma'}{ \gamma'+ \nicefrac{c_1}{c_0}\delta'} 
\label{eqn:master1} , 
\end{eqnarray}
with
\begin{eqnarray}
& \gamma' :=& \intop_{0}^{\infty}P\left(N_
{on},N_{\text{off}}|\lambda_{\text{bg}},H_{\text{0}}\right) \nonumber\\
& & \times P_{\text{0}}\left(\lambda_{\text{bg}}|H_{\text{0}}\right)d\lambda_
{\text{bg}} P_{\text{0}}\left(H_{\text{0}}\right) , \\
& \delta' :=& \intop_{0}^{\infty}\intop_{0}^{\infty}P\left(N_{on},N_{ 
off}|\lambda_{\text{s}},\lambda_{\text{bg}},H_{\text{1}}\right) \nonumber\\ 
& &\times P_{\text{0}}\left(\lambda_{\text{s}},\lambda_{\text{bg}}|H_{
\text{1}}\right)d\lambda_{\text{s}}d\lambda_{\text{bg}} P_{\text
{0}} \left(H_{\text{1}}\right) . 
\end{eqnarray}
To calculate the analytic outcome of Eqn. \ref{eqn:master1}, the 
priors for the hypothesis $P_{\text{0}}\left(H_{\text{0}}\right)$ 
and $P_{\text{0}}\left(H_{\text{1}}\right)$ have to be identified. 
Given the lack of prior information to which hypothesis is more 
likely, they are chosen to be equal
\begin{equation}
 P_{\text{0}}\left(H_{\text{0}}\right)=P_{\text{0}}\left(H_{
 \text{1}}\right)=\frac{1}{2} .
\end{equation}
When the model parameter spaces are the same, it is common to set $c_0 = c_1$. 
In the case that that the two models have differing 
dimensions, special effort has to be invested, in order to assign a value to 
$\frac{c_1}{c_0}$, based on extrinsic arguments \citep{Berger2001}. 
Therefore, imagine no counts in either region. This means no signal was observed, 
which means the signal hypothesis $H_{\text{1}}$ can not become more likely
\begin{equation}
 P\left(H_{\text{0}}|0,0\right) \geq P_{\text{0}}\left(H_{\text{0}}\right).
 \label{eqn:argue}
\end{equation}
This is a limit on the posterior model probability, that can be used 
as the basis of a robust Bayesian analysis \citep{Berger1994}. In
particular I argue that when no counts are observed, the probability 
for either model stays the same and therefore equality holds
in Eqn. 
\ref{eqn:argue}.
The approach leads to the determination of the fraction $\frac{c_1}{c_0}$ via the equation
\begin{eqnarray}
\frac{c_1}{c_0} &=& \frac{\gamma'}{\delta'}|_{N_{\text{on}},N_{\text{off}}=0} .
\label{eqn:allenextra}
\end{eqnarray}
The evaluation of Eqn. \ref{eqn:master1} together with Eqn. \ref{eqn:allenextra}
may be found in 
Appendix \ref{app:master1proof}. Altogether the probability of 
$H_{\text{0}}$ being true, given $N_{\text{on}}$ and $N_{\text{off}}$, is
\begin{equation}
P\left(H_{\text{0}}|N_{\text{on}},N_{\text{off}}\right) = 
\frac{\gamma}{\gamma+\nicefrac{c_1}{c_0} \delta} ,
 \label{eqn:ts}
\end{equation}
where $\gamma$ and $\delta$ are defined in terms of the Gamma function
$\Gamma\left(x\right)$ and the hypergeometric function
$_{2}\F_{1}\left(a,b;c;z\right)$ 
\begin{eqnarray}
\gamma & := & \left(1+2N_{\text{off}}\right) \alpha^{\nicefrac
{1}{2}+N_{\text{on}}+N_{\text{off}}} \\ &  & 
\times\Gamma\left(\nicefrac{1}{2}+N_{\text{on}}+N_{\text{off}}
\right) ,\nonumber\\ \delta & := & 2\left(1+\alpha\right)^{N_{\text
{on}}+N_{\text{off}}} \Gamma\left(1+N_{\text{on}}+N_{\text{off}} 
\right) \\ &  & \times _{2}\F_{1}\left(\nicefrac{1}{2}+N_{\text
{off}},1+N_{\text{on}}+N_{\text{off}};\nicefrac{3}{2}+N_{\text
{off}} ;-\nicefrac{1}{\alpha}\right) ,\nonumber \\
\frac{c_1}{c_0} & = & \frac{\sqrt{\pi}}{2 \arctan\left(\nicefrac{1}{\sqrt{\alpha}}\right)}.
\end{eqnarray}
Equation \ref{eqn:ts} is, however, not restricted to small 
count numbers and can, with current 
PCs and numerical tools like \textsc{Mathematica}, be easily 
calculated up to thousands of counts.

\subsection{\label{sec:det}Signal Detection}
A signal detection based on Eqn. \ref{eqn:ts} may be claimed when 
the resulting probability of the null hypothesis $H_{\text{0}}$ is 
low. In high-energy astrophysics, the consensus \citep
{li1983analysis,abdo2009fermi} p value for a source discovery is 
$p=5.7\times10^{-7}$, corresponding to a \textit{5$\sigma$ measurement}. 
Scientists frequently use lower thresholds 
for the detection of known sources. However, 
this number value is used in this paper
for the probability of $H_{\text{0}}$ 
as source detection criterion.

One must keep in mind that these are two completely
different quantities: a probability of a model, and a frequency of
an outcome. $P\left(H_{\text{0}}|N_{\text{on}},N_{\text{off}}\right)$ 
explicitly weighs alternative models, while the
frequentist result does not.

That said, and with the help of the inverse error function 
$\erf^{-1}(x)$, the Bayesian significance $S_{\text{b}}$ is 
introduced, defined as ''if the probability were normal distributed, 
it would correspond to a $S_{\text{b}}$ standard deviation 
measurement``:
\begin{equation}
S_{\text{b}}=\sqrt{2}  \erf^{-1}\left[1-P\left(H_{\text{0}}|N_{
\text{on}},N_{\text{off}}\right)\right] .
\label{eqn:sigb}
\end{equation}
Using the above equation, it is easy to compare detection 
or discovery claims with different methods and thresholds, 
as shown in Sec. \ref{sec:val}.

\subsection{\label{sec:signal}Signal Strength}
If the counted events lead to a detection, the signal parameter 
strength can be estimated. In other words, it is safe to assume 
hypothesis $H_{\text{1}}$. 
The conditional probability of the signal 
and the background parameters $\lambda_{\text{s}}$ and 
$\lambda_{\text{bg}}$ may then be calculated from Bayes' law: 
\begin{eqnarray}
\label{eqn:marg1}
 & &P\left(\lambda_{\text{s}},\lambda_{\text{bg}}|N_{\text
 {on}},N_{\text{off}},H_{\text{1}}\right) = \\ & &\frac{P\left(N_
 {\text{on}},N_{\text{off}}|\lambda_{\text{s}},\lambda_{\text
 {bg}},H_{\text{1}}\right) P_{\text{0}}\left(\lambda_{\text{s}},
 \lambda_{\text{bg}}|H_{\text{1}}\right)}{\int_{0}^{\infty}\int_
 {0}^{\infty} P\left(N_{\text{on}},N_{\text{off}}|\lambda_{\text
 {s}},\lambda_{\text{bg}},H_{\text{1}}\right) P_{\text{0}}
 \left(\lambda_{\text{s}},\lambda_{\text{bg}}|H_{\text{1}}
 \right)d\lambda_{\text{s}}d\lambda_{\text{bg}}}. \nonumber
\end{eqnarray}
Given the data, one would like to infer the signal
$\lambda_{\text{s}}$ without reference to $\lambda_{\text{bg}}$ but
fully accounting for the uncertainty on $\lambda_{\text{bg}}$.  This
can be done by marginalizing over the nuisance parameter
$\lambda_{\text{bg}}$:
\begin{equation}
 P\left(\lambda_{\text{s}}|N_{\text{on}},N_{\text{off}},H_{\text
 {1}}\right)=\int_{0}^{\infty}P\left(\lambda_{\text{s}}, 
 \lambda_{\text{bg}}|N_{\text{on}},N_{\text{off}},H_{\text{1}}
 \right)d\lambda_{\text{bg}} .
\label{eqn:marg2}
\end{equation}
Equation \ref{eqn:marg2} can be analytically calculated using Eqn. 
\ref{eqn:likelihood}, \ref{eqn:prior}, and the results from 
Appendix \ref{app:master1proof}, which is done in Appendix \ref{app:sigstr}.
The improper prior is acceptable because the proportionality constant 
$c_1$ cancels and the posterior is proper. 
The result may be expressed in terms of three functions, 
namely the Poisson distribution 
$P_{\text{P}}\left(N|\lambda\right)$, the regularized hypergeometric 
function $_{2}\tilde{\F}_{1}\left(a,b;c;z\right) = 
\nicefrac{_{2}\F_{1}\left(a,b;c;z\right)}{\Gamma\left(c\right)}$, and 
the Tricomi confluent hypergeometric function $\U\left(a,b,z\right)$:
\begin{eqnarray}
 & & P\left(\lambda_{\text{s}}|N_{\text{on}},N_{\text{off}},H_{
 \text{1}}\right) = P_{\text{P}}\left(N_{\text{on}}+N_{\text
 {off}}|\lambda_{\text{s}}\right)  \label{eqn:posterior}\\ & & 
 \times \frac{\U\left[\nicefrac{1}{2}+N_{\text{off}},1+N_{\text
 {off}}+N_{\text{on}},\left(1+\nicefrac{1}{ \alpha}
 \right)\lambda_{\text{s}}\right]}{_{2}\tilde{\F}_{1}
 \left(\nicefrac{1}{2}+N_{\text{off}} ,1+N_{\text{off}}+N_{\text
 {on}};\nicefrac{3}{2}+N_{\text{off}};-\nicefrac{1}{\alpha}
 \right)} . \nonumber
\end{eqnarray}
This posterior contains the full information. In order to quote 
numbers one may take the mode $\lambda_{\text{s}}^*$,
which is the value of $\lambda_{\text{s}}$ that maximizes the 
posterior distribution 
$P\left(\lambda_{\text{s}}|N_{\text{on}},N_{\text{off}},H_{\text{1}}
\right)$, as signal estimator. The error on the quoted signal can be 
evaluated from the cumulative distribution function. For instance to 
get the smallest Bayesian interval (also known as Highest Posterior Density 
interval or HPD) containing the signal parameter 
with $68\%$ probability one can solve 
\begin{equation}
 0.68 = \int_{\lambda_{\text{min}}}^{\lambda_{\text{max}}}P
 \left(\lambda_{\text{s}}|N_{\text{on}},N_{\text{off}},H_{\text
 {1}}\right)d\lambda_{\text{s}} ,
\end{equation}
together with the constraint
\begin{equation}
 P\left(\lambda_{\text{min}}|N_{\text{on}},N_{\text{off}},H_{
 \text{1}}\right) = P\left(\lambda_{\text{max}}|N_{\text{on}},N_{
 \text{off}},H_{\text{1}}\right) ,
\end{equation}
numerically for $\lambda_{\text{min}}$ and $\lambda_{\text{max}}$.
The final result may be quoted as
\begin{equation}
 \lambda_{\text{s}}=\lambda_{\text{s}-(\lambda_{\text{s}}^*-
 \lambda_{\text{min}})}^{*+(\lambda_{\text{max}}-\lambda_{\text
 {s}}^*)}.
\end{equation}

\subsection{\label{sec:ul}Signal Upper Limit}
If the data show no significant detection, an upper limit on the 
signal parameter may be calculated, assuming that the signal is 
there (i.e. $H_{\text{1}}$ is true) but too weak to be measured. For 
example a $99\%$ probability limit $\lambda_{99}$ on the signal 
parameter $\lambda_{\text{s}}$ is calculated by solving
\begin{equation}
 \intop_{0}^{\lambda_{99}}P\left(\lambda_{\text{s}}|N_{\text
 {on}},N_{\text{off}},H_{\text{1}}\right)d\lambda_{\text{s}} 
 =0.99 .
\end{equation}
This result comes naturally in a Bayesian approach of the problem 
but is hard to calculate in a frequentist approach. In particular 
frequentists struggle with the marginalization of the problem and 
with special cases at the border of the parameter space, all of 
which lead to ad hoc adjustments without theoretical justification 
\citep{rolke2005limits}. The only practical remedy comes from Monte 
Carlo studies that in fact such limits with adjustments have (at 
least) the claimed frequentist coverage. In this Bayesian approach, 
all possible values in the parameter space are dealt with in a 
uniform way, no matter if zero counts or thousands of counts. The 
signal upper limit result is in particular interesting for 
$N_{\text{on}}=N_{\text{off}}=0$. It underlines that measuring zero is
different from not measuring at all, hence valid limits can be
derived. Importantly the estimates are always physically meaningful
(i.e. positive
$\lambda_{\text{s}}^{*},\lambda_{min},\lambda_{max},\lambda_{99},...$
).

\section{\label{sec:val}Validation}
Jeffreys's prior is constructed by a formal rule \citep{jeffreys1961theory}
and motivated by the requirement for invariance under one-to-one transforms. 
However this is not the only possible choice and, 
when data are sparse, the choice of the prior is important.
In order to validate that it is a reasonable choice 
I compare it to the prior from \citet{gregory2005bayesian},
the frequentist solution from \citet{li1983analysis},
and to a simulation.

\subsection{\label{sec:comp}Model Comparison}
For the On/Off problem one alternative with informative flat priors was presented by \citet{gregory2005bayesian}.
The hypothesis test is, in this case, dependent on the prior signal upper boundary $\lambda_{\textrm{smax}}$ 
in addition to $N_{\textrm{on}}$, $N_{\textrm{off}}$ and $\alpha$.
Therefore reasonable assumptions on the signal upper boundary $\lambda_{\textrm{smax}}$ have to be made, in order to 
compare Eqn. 14.24, \citet{gregory2005bayesian} with Eqn. \ref{eqn:ts}. 
The signal posteriors however can be compared directly as they depend only on the three initial parameters
in both cases. 
\begin{figure}[ht]
\centering
\includegraphics[width=0.49\textwidth]
{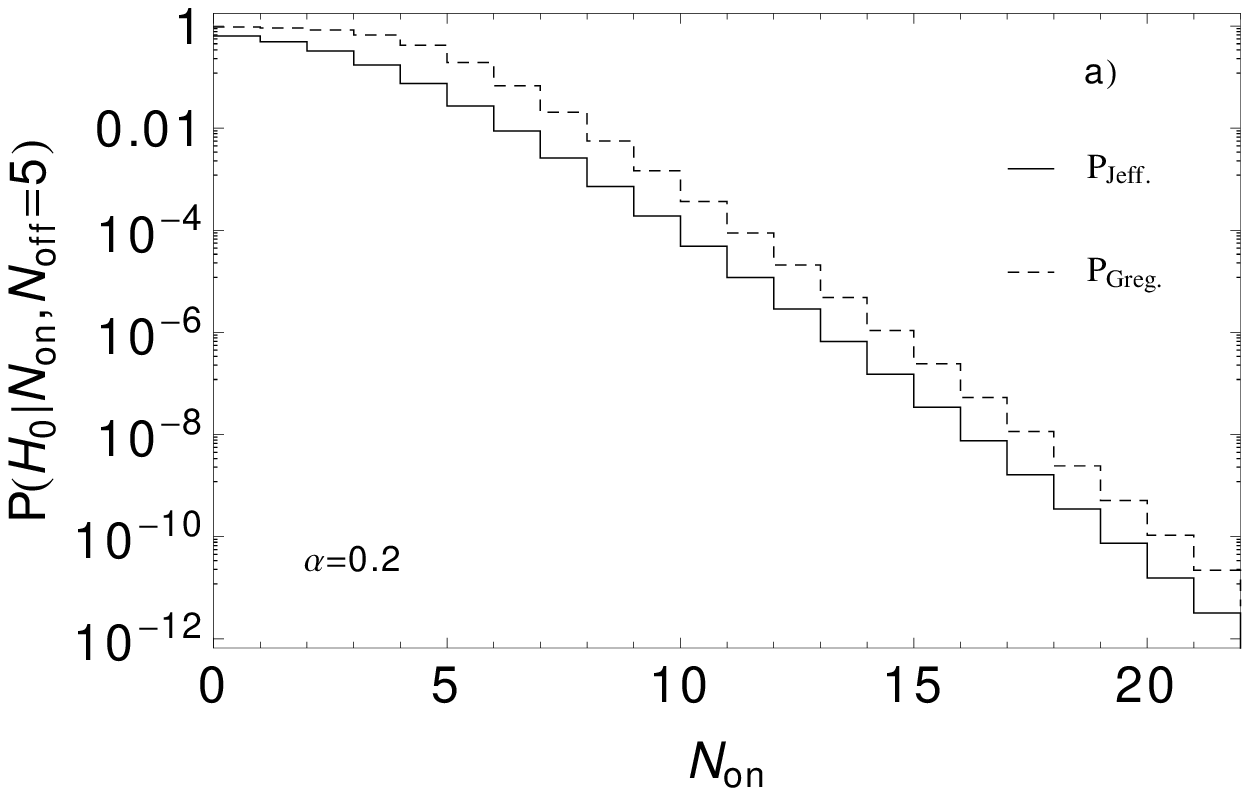}
\includegraphics[width=0.49\textwidth]
{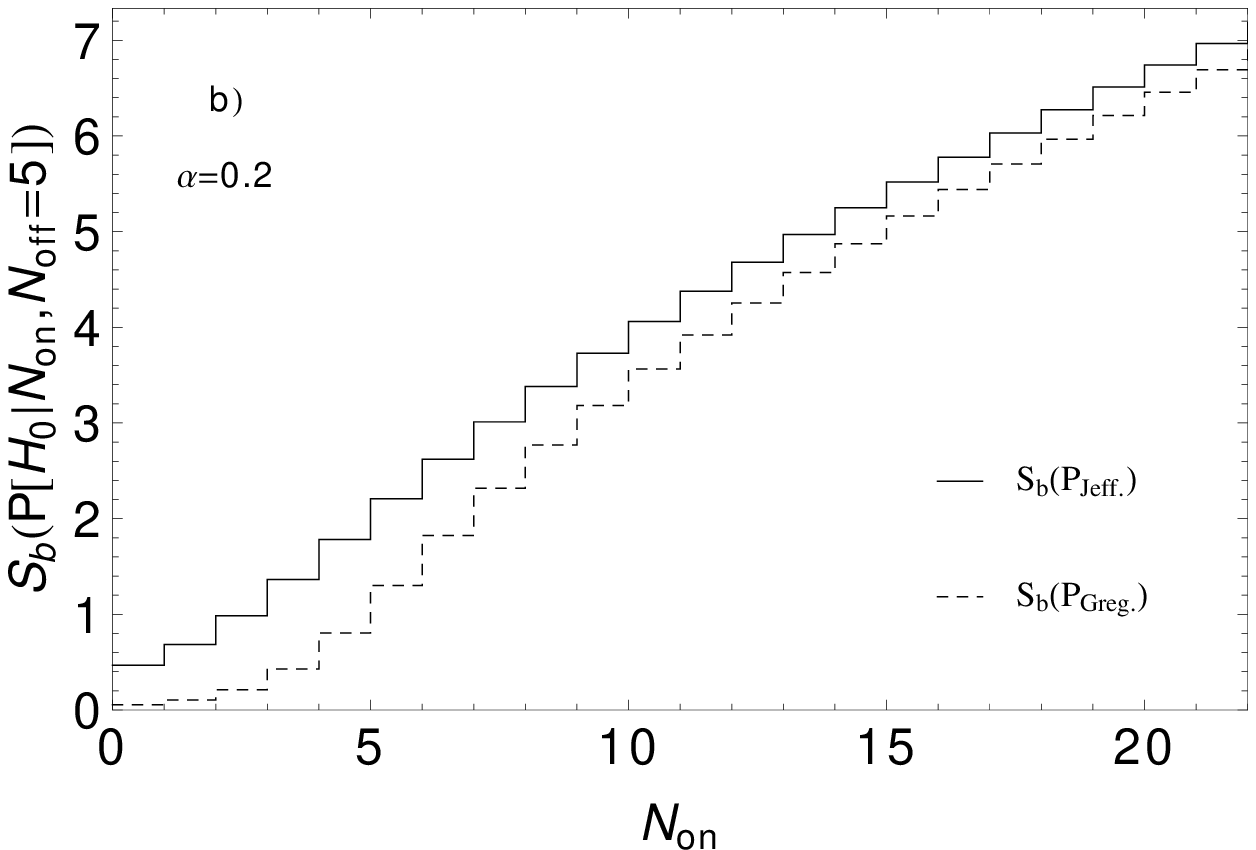}\\
\includegraphics[width=0.49\textwidth]
{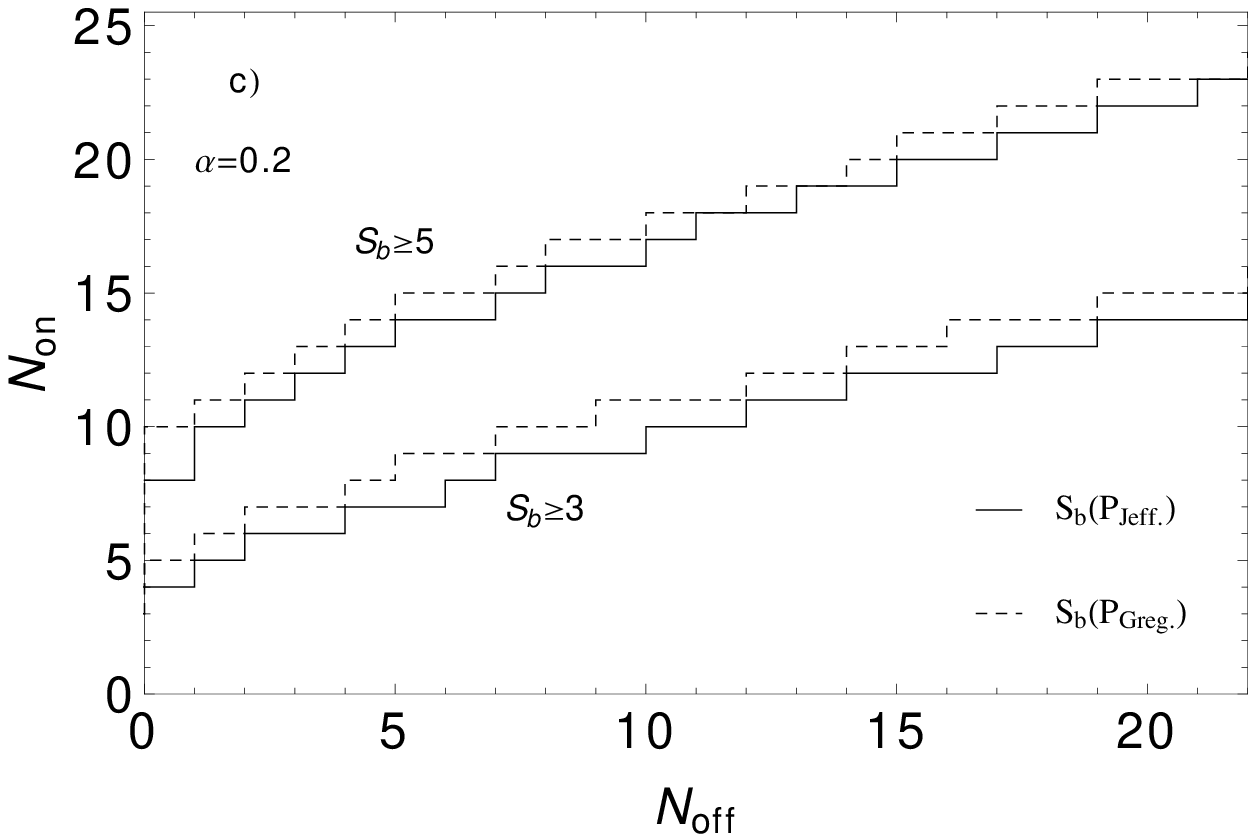}
\includegraphics[width=0.49\textwidth]
{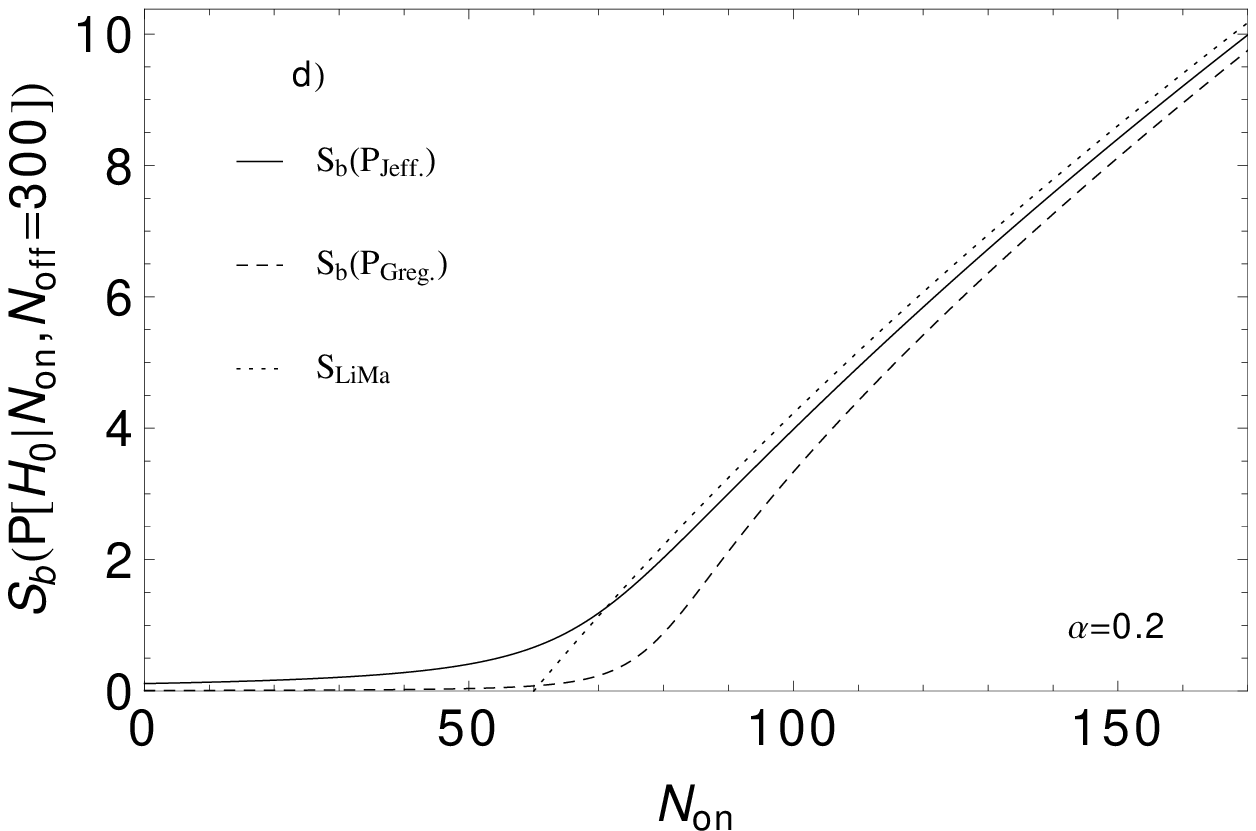}\\
\caption{\label{fig:comp1}Comparison of the On/Off hypothesis test with Jeffreys's and Gregory's priors for low and 
large count numbers. In a) the null hypothesis posterior probability is shown as a function of $N_{\text{on}}$
in the low counts regime. b) shows the same but using the nonlinear Bayesian significance scale Eqn. \ref{eqn:sigb}.
In c) the limiting curves for which $N_{\text{on}}: S_{\text{b}} \geq 3$ or $S_{\text{b}} \geq 5$ are shown. d) shows a comparison 
in the large counts regime and additionally \citet{li1983analysis}, Eqn. 17.}
\end{figure}  
\begin{figure}[ht]
\centering
\includegraphics[width=0.49\textwidth]
{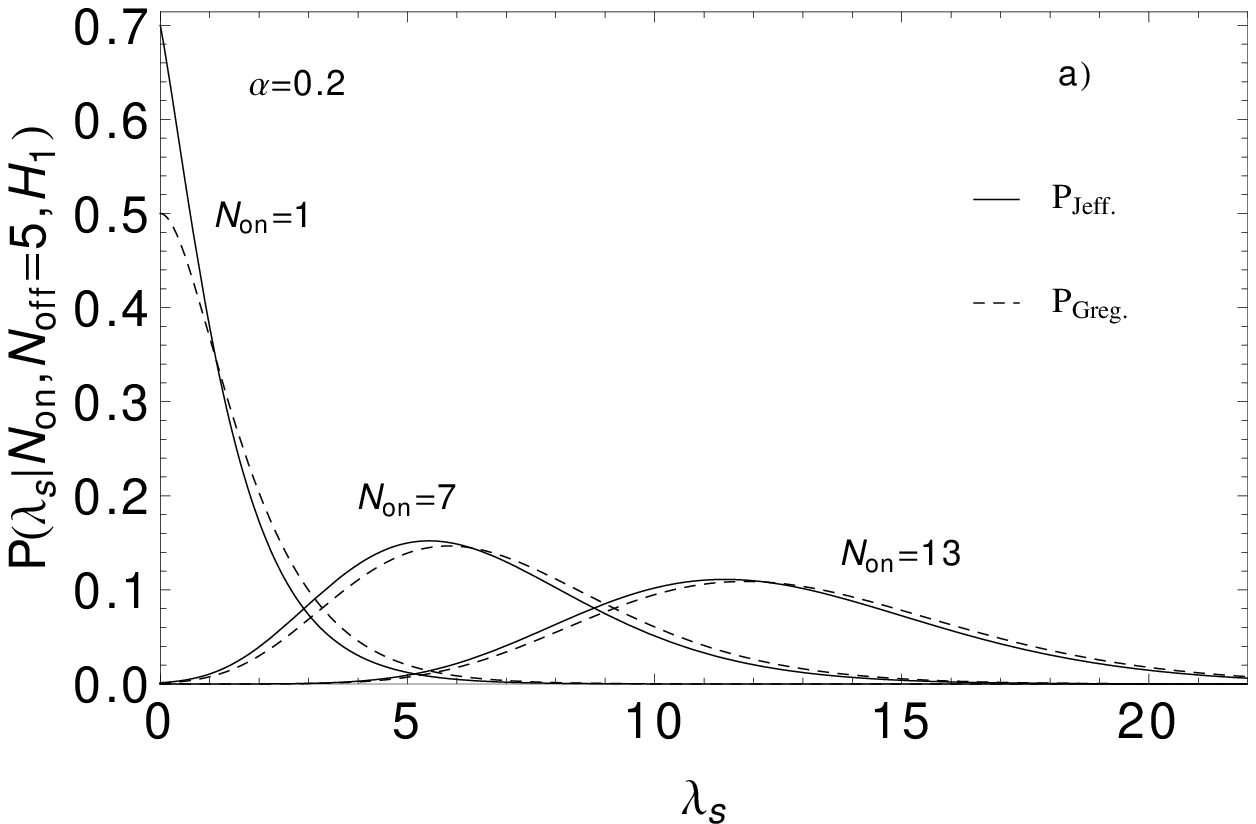}
\includegraphics[width=0.49\textwidth]
{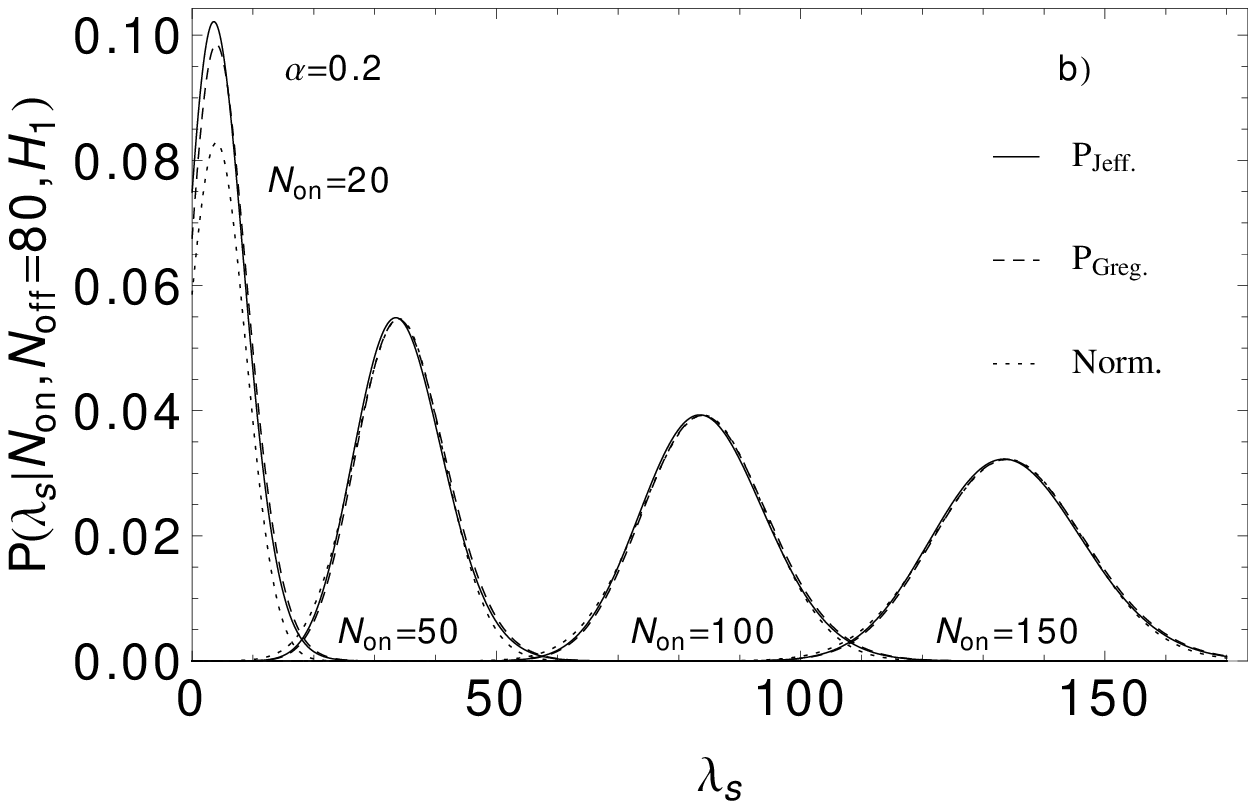}\\
\caption{\label{fig:comp2}Comparison of the signal posteriors with Jeffreys's and Gregory's priors. 
a) shows a comparison for low count numbers. b) shows a comparison for high count numbers and 
shows additionally a normal distribution with mean Eqn. \ref{eqn:direct} and variance Eqn. 
\ref{eqn:direct2}}
\end{figure}  

The hypothesis test comparison is shown in Fig. \ref{fig:comp1}. 
Fig. \ref{fig:comp1}a, \ref{fig:comp1}b, and \ref{fig:comp1}c show the situation for a typical low count case
with $\alpha=0.2$ and an assumed $\lambda_{\textrm{smax}}=22$ such that 
a signal detection with that strength would be without any doubt. 
Gregory's prior shows a similar behavior as the Jeffreys's prior, 
but is slightly shifted towards higher probabilities for the null hypothesis
$P\left(H_{\text{0}}|N_{\text{on}},N_{\text{off}}\right)$ or lower significance
$S_{\textrm{b}}$ (Eqn. \ref{eqn:sigb}).
In Fig. \ref{fig:comp1}c one can see the limiting curve for which $N_{\textrm{on}}$, given $N_{\textrm{off}}$,
the significance $S_{\textrm{b}}$ is $\geq3$ or $\geq5$.
This shows that, when it comes to decision making in the low-count regime, both models are 
mostly within one count from another. 

Figure \ref{fig:comp1}d shows a comparison of the different priors for high count numbers 
with $\alpha=0.2$, $N_{\textrm{off}}=300$ and $\lambda_{\textrm{smax}}=170$. 
Additionally, the methods are compared to the frequentist result of \citet{li1983analysis}, Eqn. 17.  
Both methods appear to converge 
on Li \& Ma's result for large count numbers, but Jeffreys's prior gives 
a closer approximation.
From a physical point of view one could argue that for $N_{\textrm{on}} \sim \alpha N_{\textrm{on}}$,
$P\left(H_{\text{0}}|N_{\text{on}},N_{\text{off}}\right)$ should be close to the prior 
model probability $P_0\left(H_{\text{0}}\right) = 0.5$
because any one $N_{\textrm{on}}$ more should increase the probability for the signal hypothesis $H_1$ 
and any $N_{\textrm{on}}$ less should decrease it. Numerically it seems that this is the case 
for Jeffreys's prior, but not for Gregory's prior.

When comparing the signal posterior, Jeffrey's and Gregory's priors agree well. Fig. \ref{fig:comp2}a
shows a comparison of the two methods for low count numbers with $\alpha=0.2$, $N_{\textrm{off}}=5$.
The differences in the signal posterior are marginal.
For high count numbers, as shown in Fig. \ref{fig:comp2}b with $\alpha=0.2$, $N_{\textrm{off}}=80$, 
both methods converge quickly to the 
to the classical result
of a normal distribution with 
\begin{eqnarray}
  & E(\lambda_{\textrm{s}}) & = N_{\textrm{on}} - \alpha N_{\textrm{off}} \label{eqn:direct}\\
  & Var(\lambda_{\textrm{s}}) & = N_{\textrm{on}} + \alpha^2 N_{\textrm{off}} \label{eqn:direct2}.  
\end{eqnarray}
However, because of the subtraction and when $N_{\textrm{on}} \sim \alpha N_{\textrm{on}}$ the normal distribution 
can include negative values for the signal confidence region. 
This problem is resolved using the Bayesian methods.

Overall the results are comparable and show that the results obtained by Jeffreys's prior are sensible. 
They behave well in all test-case examples, in particular at $N_{\textrm{on}} \sim \alpha N_{\textrm{on}}$, 
and converge to the other results for high count numbers.

\subsection{\label{sec:sim}Simulations}
To further verify the method developed in this paper, the hypothesis test and the maximum signal posterior are calculated 
for a simulated set of observations.
First, one thousand $N_{\textrm{on}}$ and $N_{\textrm{off}}$ are drawn randomly from two Poisson distributions with 
means of Eqn. \ref{eqn:mean1} and Eqn. \ref{eqn:mean2}. 
Second, the developed methods are applied to the simulation, as well as Li \& Ma's method for the hypothesis test and 
a direct background subtraction (Eqn. \ref{eqn:direct}) for the signal strength. 
Then, the results of these methods are compared to one another and to the true
parameters. These are $\lambda_{\textrm{bg}} = 300$ for the background strength and
$\lambda_{\textrm{s}} = 50$ for the signal strength. The ratio of exposure $\alpha$ is 1/5.
\begin{figure}[ht]
\centering
\includegraphics[width=0.49\textwidth]
{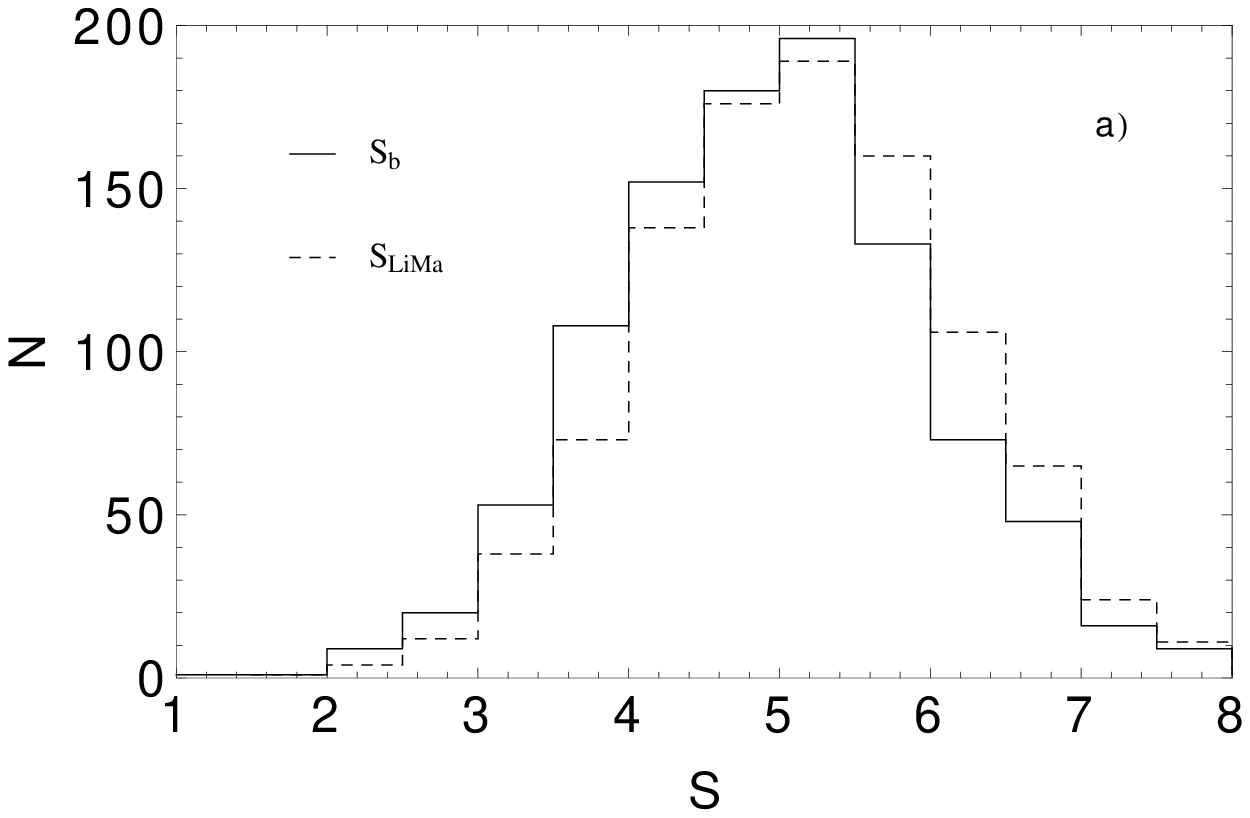}
\includegraphics[width=0.49\textwidth]
{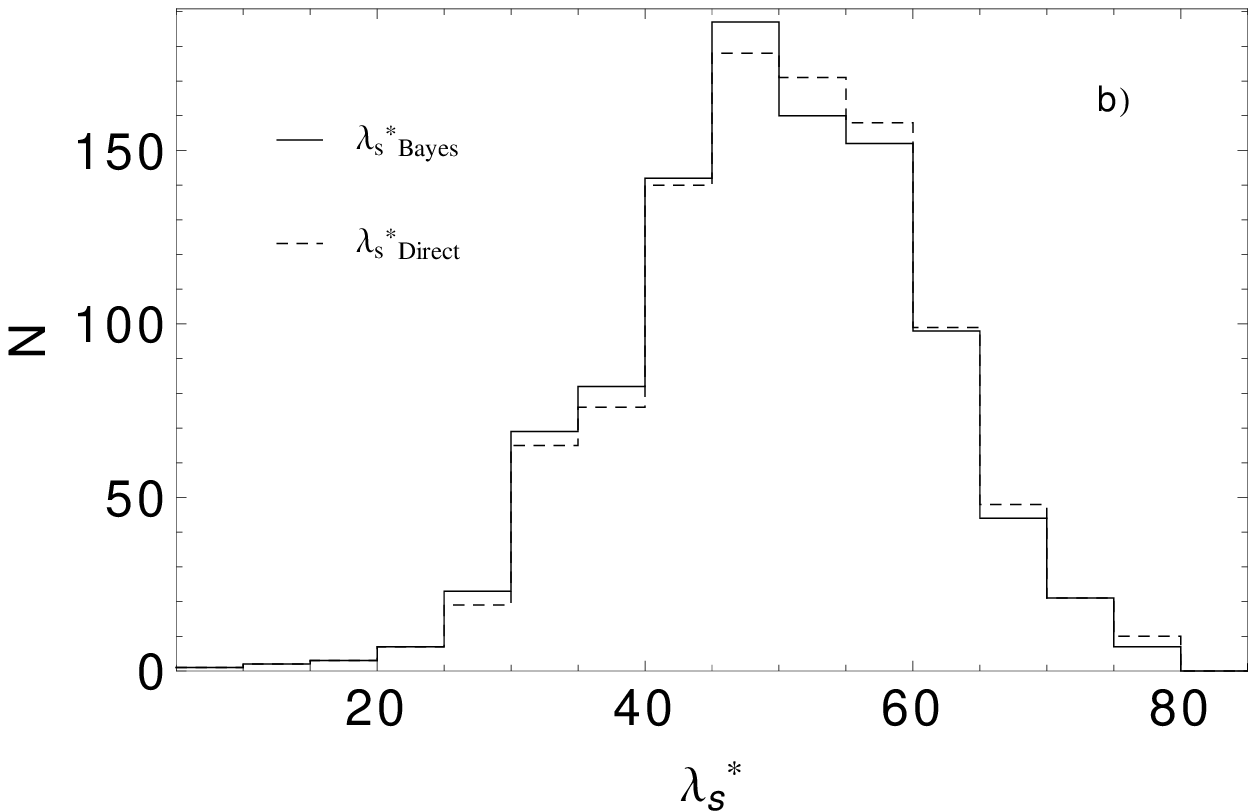}\\
\caption{\label{fig:sim}Simulation results of the analysis, in comparison to Li \& Ma and 
direct background subtraction.}
\end{figure}

Figure \ref{fig:sim}a shows the hypothesis test simulation results. 
Li \& Ma's test statistic shows a small systematic shift in comparison to the Bayesian significance 
(Eqn. \ref{eqn:sigb}) in agreement with the results from Sec. \ref{sec:comp}. 
In Fig. \ref{fig:sim}b the signal strength comparison is shown. 
The parameter $\lambda_{\text{s}}^*$ is first calculated 
as the maximum of the marginalized posterior (Eqn. \ref{eqn:posterior})
and second with direct background subtraction. Both methods agree well and 
can reconstruct the true signal parameter $\lambda_{\textrm{s}} = 50$
with similar errors.

\section{\label{sec:grb}Application: Gamma-Ray Bursts} Gamma-ray 
bursts (GRB) are extraterrestrial flashes of gamma-rays, lasting only few 
seconds mostly. One interesting question is whether gamma-ray bursts 
produce very high energy ($>100$GeV) gamma-rays, as proposed by 
some theories \citep[e.g.][]{abdo2009fermi}. Because of their duration and 
fluence,
gamma-ray satellites and Cherenkov telescopes measure only few 
events during the flare itself, or shortly thereafter. In Tab. \ref
{tab:data} data from 12 gamma-ray burst observations, made by the 
space based
\textsc{Fermi} Large Area Telescope (\mbox{\textsc{Fermi}-LAT}) 
 and the ground based \mbox{\textsc{VERITAS}} Cherenkov telescope, are compiled.
\footnote{The $\alpha$ parameter of the 
\mbox{\textsc{VERITAS}} observations is calculated with 
their employed test statistic \textit{ratio of Poisson means},
according to
\citet{cousins2008evaluation}. It is solved numerically for $\alpha$, 
except for the case of GRB 080330 where not a single count was 
measured in the \textit{on} region and no significance is given. In 
this case, the $\alpha$ parameter is assumed to be the mean of the 
other \mbox{\textsc{VERITAS}} wobble-mode observations.}
\begin{figure}
\centering
\includegraphics[width=0.47\textwidth]
{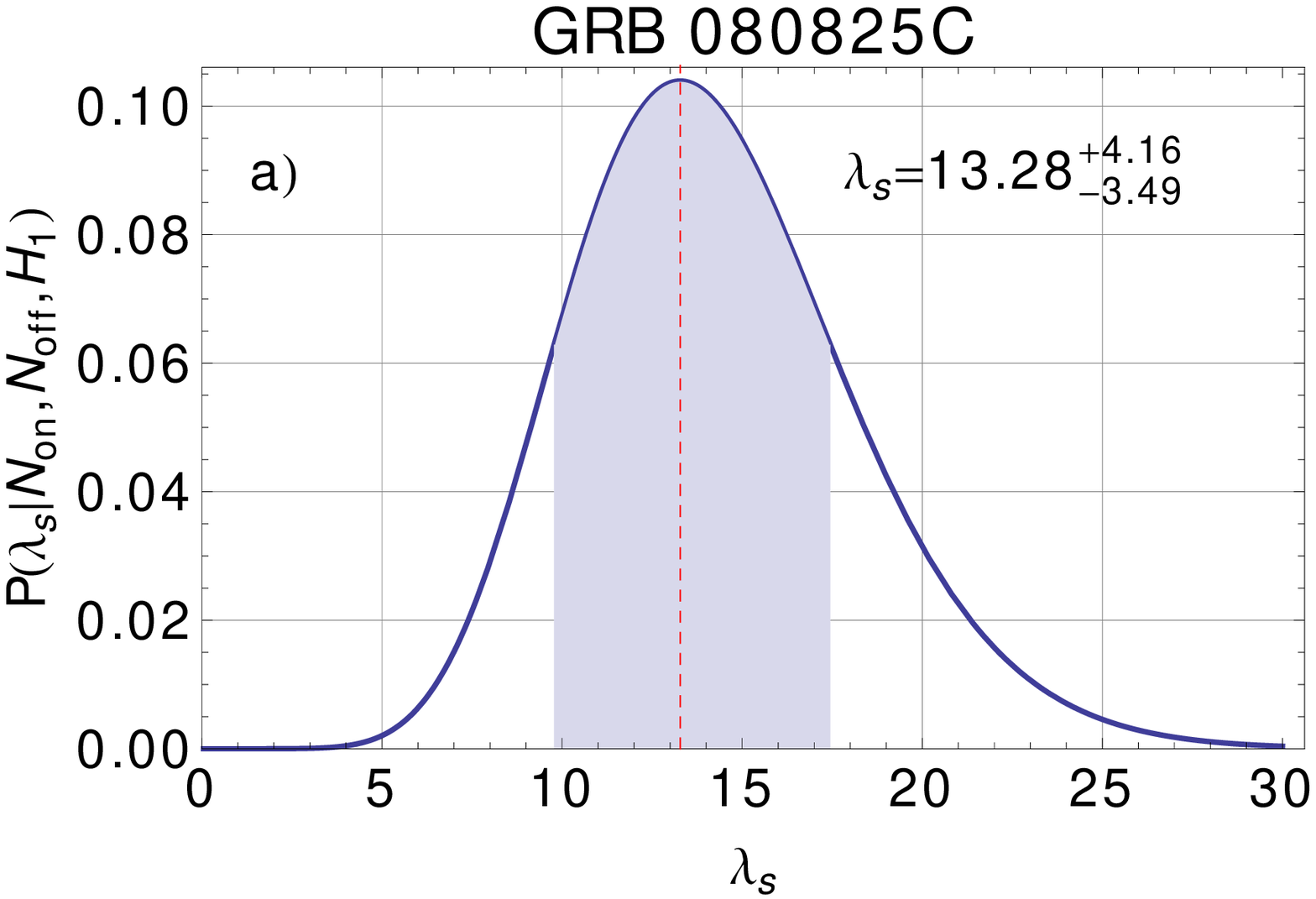}
\includegraphics[width=0.47\textwidth]
{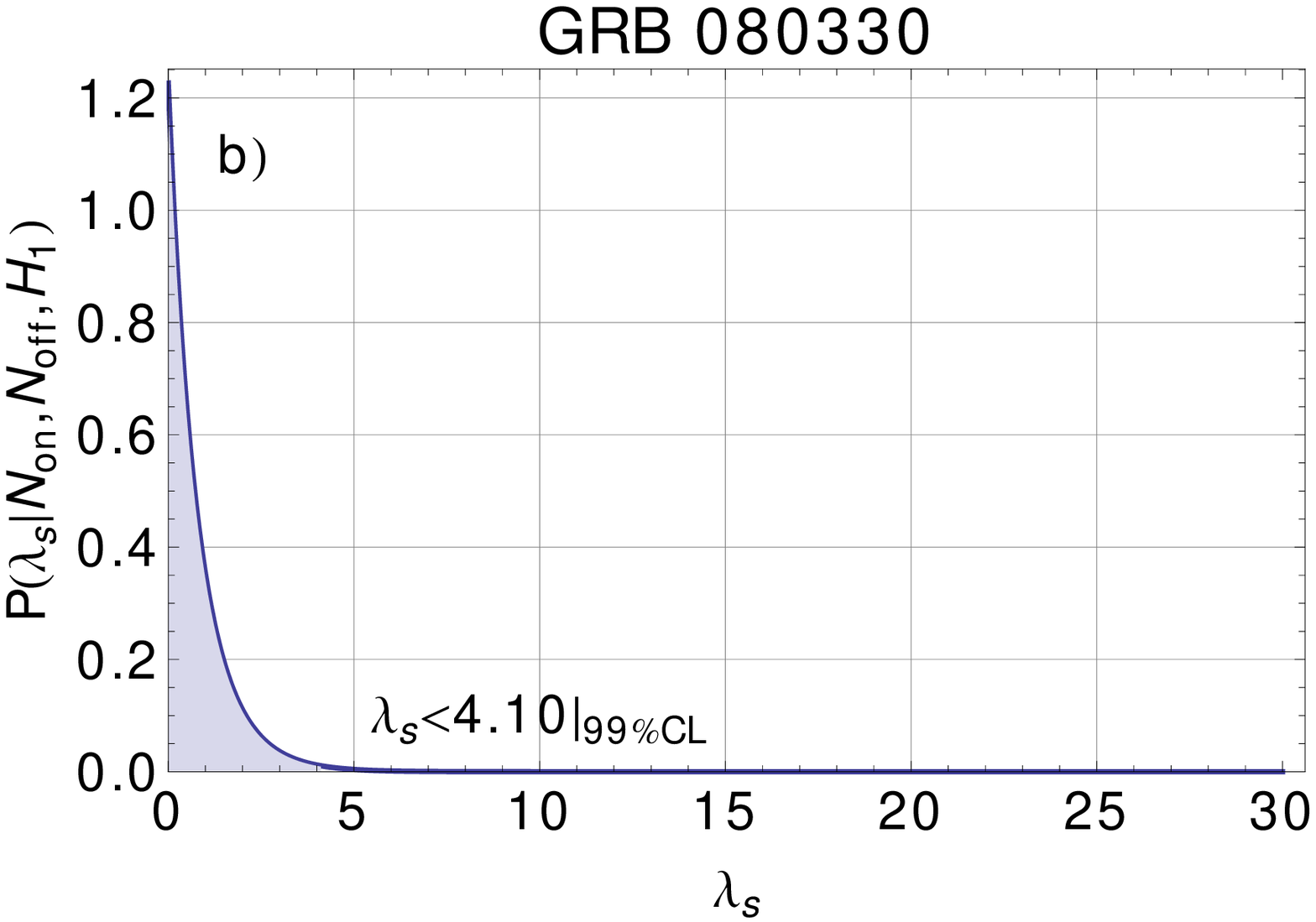}\\
\caption{\label{fig:pdfs}Comparison of the marginal signal parameter 
$\lambda_{\text{s}}$ posterior distribution for two gamma-ray 
bursts. In a), GRB 080825C, a source is detected and 
the signal parameter estimate and its smallest $68\%$ credibility 
interval is calculated. In b) an upper limit is 
calculated for GRB 080330 where no \textit{on} event was measured.}
\end{figure}




\begin{deluxetable}{ccccccccccccc}
\tabletypesize{\scriptsize}



\tablecaption{Low count gamma-ray burst data\label{tab:data}}

\tablenum{1}

\tablehead{\colhead{GRB}  & \colhead{$N_{\text{on}}$} & \colhead{$N_{\text{off}}$} & \colhead{$\alpha$ } & \colhead{$S_{\text{lm}}$}  & \colhead{ $P(H_{\text{0}}|N_{\text{on}},N_{\text{off}})$} & \colhead{ $S_{\text{b}}$} & \colhead{$\lambda_{\text{s}}$} & \colhead{$\lambda_{99}$} &\colhead{$\lambda_{\textrm{Ref.}}$} &\colhead{$\lambda_{99}^{\textrm{Rolke}}$} & \colhead{Reference} } 

\startdata
 070419A  & 2            & 14            & \textbf{0.057} &   \ldots               & \textbf{0.28}           & \textbf{1.09}        &  \ldots                & \textbf{6.88} &  \ldots & \textbf{7.34}	& 1	 \\ 
 070521  & 3            & 113           & \textbf{0.057} &  \ldots                & \textbf{0.72}           & \textbf{0.36}        &  \ldots                & \textbf{6.12} &  \ldots & \textbf{3.52} 	& 1	 \\ 
 070612B & 3            & 21            & \textbf{0.066} &  \ldots                & \textbf{0.26}           & \textbf{1.12}        &  \ldots                & \textbf{8.00} &  \ldots & \textbf{8.54}	 & 1 	 \\ 
 080310  & 3            & 23            & \textbf{0.128} &  \ldots                & \textbf{0.51}           & \textbf{0.66}        &  \ldots                & \textbf{7.16} &  \ldots & \textbf{7.08}	 & 1	 \\ 
 080330  & 0            & 15            & \textbf{0.123} &   \ldots               & \textbf{0.70}           & \textbf{0.39}        &  \ldots                & \textbf{4.10} &  \ldots & \textbf{2.40}	 & 1	 \\ 
 080604  & 2            & 40            & \textbf{0.063} &   \ldots               & \textbf{0.58}           & \textbf{0.55}        &  \ldots                & \textbf{6.12} &  \ldots & \textbf{5.66}	 & 1  \\ 
 080607  & 4            & 16            & \textbf{0.112} &   \ldots               & \textbf{0.21}           & \textbf{1.27}        &  \ldots                & \textbf{9.17} &  \ldots & \textbf{9.83}	 & 1	\\ 
 080825C & 15           & 19            & 0.063          & 6.4            & \textbf{9.66E-10}       & \textbf{6.11}        & \textbf{$\mathbf{13.28_{-3.49}^{+4.16}}$}   & \ldots &  13.7    &  \ldots             &		2	 \\ 
 081024A & 1            & 7             & \textbf{0.142} &  \ldots                & \textbf{0.50}           & \textbf{0.67}        &  \ldots                & \textbf{5.29} &  \ldots  & \textbf{5.19}	& 1	 \\ 
 090418A & 3            & 16            & \textbf{0.123} &  \ldots                & \textbf{0.37}           & \textbf{0.89}        &   \ldots               & \textbf{7.64} &  \ldots  & \textbf{8.01}	& 1 	\\ 
 090429B & 2            & 7             & \textbf{0.106} &  \ldots                & \textbf{0.26}           & \textbf{1.13}        &   \ldots               & \textbf{6.92} &  \ldots  & \textbf{7.41}	& 1 	\\ 
 090515  & 4            & 24            & \textbf{0.126} &  \ldots               & \textbf{0.41}           & \textbf{0.83}        &   \ldots               & \textbf{8.34}  &  \ldots & \textbf{8.66}	& 1 	\\ 
\enddata

\tablecomments{Table \ref{tab:data} shows low count gamma-ray burst
data where either or both of $N_{\text{on}}$ 
or $N_{\text{off}}$ are 
$\leq15$. Only GRB 080825C was detected with a high significance. Its 
reference source strength $\lambda_{\text{Ref.}}$ is given in \citet{abdo2009fermi}. All
calculated values are in bold. The ratio of exposure $\alpha$ is calculated according to 
\citet{acciari2011veritas}. The $\alpha$ values are then used for calculating Rolke's upper limit
$\lambda_{99}^{\textrm{Rolke}}$. 
}
\tablerefs{(1) \citet{acciari2011veritas}, (2) \citet{abdo2009fermi}}

\end{deluxetable}
Due to the difficulty of the detection, reporting of those events is usually an upper limit 
with low statistics. Only for the \mbox{\textsc{Fermi}-LAT}
GRB 080825C there was significant evidence to report a 
discovery. For this gamma-ray burst, the probability of the 
background-only model $P(H_{\text{0}}|N_{\text{on}},N_{\text{off}})$ 
is \mbox{$9.66\times10^{-10}$} and the gamma-ray burst is therefore 
detected. The significance expressed on the nonlinear scale (Eqn. 
\ref{eqn:sigb}) is \mbox{ $S_{\text{b}}=6.11$}. This is comparable 
to the Li \& Ma result of \mbox{$S_{\text{lm}}=6.4$}. 

In the second step, the most likely value of the signal parameter 
and the smallest $68\%$ credibility interval is calculated. The 
method is demonstrated in Fig. \ref{fig:pdfs}. The result is that 
$\lambda_{\text{s}} = 13.28_{-3.49}^{+4.16}$, which is in good agreement
to the published reference of $\lambda_{\text{Ref.}}=13.7$. 

In the same figure an 
observation from the gamma-ray burst GRB 080330, for which discovery 
cannot be claimed, is shown. For GRB 080330 and all other gamma-ray 
bursts, the data do not show evidence for a source of gamma-rays. In 
this case, upper limits $\lambda_{99}$ are calculated. All results 
are summarized in Table \ref{tab:data} and are compared 
to the Rolke $99\%$ upper limits $\lambda_{99}^{\textrm{Rolke}}$, which VERITAS used.
GRB 080330 is special in the 
sense that not even one \textit{on} event was measured. 
The results are mostly in good agreement
but, especially at the border of the parameter space 
for $N_{\text{on}} \leq \alpha N_{\text{off}}$, also deviate from another and show the 
limit of Rolke's method - which are overcome by the Bayesian method.

\section{\label{sec:con}Conclusion}
Many particle physicists, cosmic-ray physicists and high-energy astrophysicists 
struggle with sparse On/Off data. With this new Bayesian method, 
that overcomes the weaknesses of the presently used methods, it is
possible to go down to single count On/Off measurements. 
Claiming detections, setting credibility intervals, or setting upper limits is
unified in a single and consistent method.

\acknowledgments

I would like to thank F. Beaujean, A. Biland, A. Caldwell, O. Grimm, and G. 
Hughes for the inspiring discussions and invaluable comments on the 
article.






\appendix

\section{\label{app:jeffrey}Jeffreys's Priors}
Calculation of the prior of $\lambda_{\text{bg}}$ in the 
$H_{\text{0}}$ model:
\begin{eqnarray}
 &  & P_{\text{0}}\left(\lambda_{\text{bg}}|H_{\text{0}}
 \right)=\boldsymbol{(}-\sum_{N_{\text{on}}=0}^{\infty}\sum_{N_{
 \text{off}}=0}^{\infty}\nonumber\\ &  & \left\{\partial_{
 \lambda_{\text{bg}}}^2\ln\left[P_{\text{P}}\left(N_{\text{on}}|
 \alpha\lambda_{\text{bg}}\right)  P_{\text{P}}\left(N_{\text
 {off}}|\lambda_{\text{bg}}\right)\right]\right\} \nonumber\\ 
 &  & \times P_{\text{P}}\left(N_{\text{on}}|\alpha\lambda_{
 \text{bg}}\right)  P_{\text{P}}\left(N_{\text{off}}|\lambda_{
 \text{bg}}\right)\boldsymbol{)}^{\frac{1}{2}}= \sqrt{\frac{1+
 \alpha}{\lambda_{\text{bg}}}}. \nonumber\\
\end{eqnarray}
Calculation of the prior for $\lambda_{\text{s}}$ and 
$\lambda_{\text{bg}}$ in the $H_{\text{1}}$ model:
\begin{eqnarray}
 &  & I_{s,s}=-\sum_{N_{\text{on}}=0}^{\infty}\sum_{N_{\text
 {off}}=0}^{\infty}\nonumber\\ &  & \left\{\partial_{\lambda_{
 \text{s}}}^2\ln\left[P_{\text{P}}\left(N_{\text{on}}|\lambda_{
 \text{s}}+\alpha\lambda_{\text{bg}}\right)  P_{\text{P}}
 \left(N_{\text{off}}|\lambda_{\text{bg}}\right)\right]\right\}
  \nonumber\\ &  & \times P_{\text{P}}\left(N_{\text{on}}|
 \lambda_{\text{s}}+\alpha\lambda_{\text{bg}}\right)  P_{\text
 {P}}\left(N_{\text{off}}|\lambda_{\text{bg}}
 \right)=\nonumber\\ &  & \frac{1}{\alpha\lambda_{\text{bg}}+
 \lambda_{\text{s}}},
\end{eqnarray}
\begin{eqnarray}
 &  & I_{bg,bg}=-\sum_{N_{\text{on}}=0}^{\infty}\sum_{N_{\text
 {off}}=0}^{\infty}\nonumber\\ &  & \left\{\partial_{\lambda_{
 \text{bg}}}^2\ln\left[P_{\text{P}}\left(N_{\text{on}}|\lambda_{
 \text{s}}+\alpha\lambda_{\text{bg}}\right)  P_{\text{P}}
 \left(N_{\text{off}}|\lambda_{\text{bg}}\right)\right]\right\}
  \nonumber\\ &  & \times P_{\text{P}}\left(N_{\text{on}}|
 \lambda_{\text{s}}+\alpha\lambda_{\text{bg}}\right)  P_{\text
 {P}}\left(N_{\text{off}}|\lambda_{\text{bg}}
 \right)=\nonumber\\ &  & \frac{\alpha\lambda_{\text{bg}}+
 \alpha^{2}\lambda_{\text{bg}}+\lambda_{\text{s}}}{
 \lambda_{\text{bg}}(\alpha\lambda_{\text{bg}}+\lambda_{\text{s}})},
\end{eqnarray}
\begin{eqnarray} 
 &  & I_{s,bg}=I_{bg,s}=-\sum_{N_{\text{on}}=0}^{\infty}\sum_{N_{
 \text{off}}=0}^{\infty}\nonumber\\ &  & \left\{\partial_{
 \lambda_{\text{bg}}}\partial_{\lambda_{\text{s}}}\ln\left[P_{
 \text{P}}\left(N_{\text{on}}|\lambda_{\text{s}}+\alpha\lambda_
 {\text{bg}}\right)  P_{\text{P}}\left(N_{\text{off}}|\lambda_{
 \text{bg}}\right)\right]\right\} \nonumber\\ &  & \times P_{
 \text{P}}\left(N_{\text{on}}|\lambda_{\text{s}}+\alpha\lambda_
 {\text{bg}}\right)  P_{\text{P}}\left(N_{\text{off}}|\lambda_{
 \text{bg}}\right)= \nonumber \\ &  & \frac{\alpha}{
 \alpha\lambda_{\text{bg}}+\lambda_{\text{s}}}.
\end{eqnarray}
The off-diagonal elements are equal, because the matrix is symmetric 
(symmetry of second derivatives). The final result for 
$P_{\text{0}}\left(\lambda_{\text{s}},\lambda_{\text{bg}}|H_{\text{1}}
\right)$ is
\begin{eqnarray}
 & & P_{\text{0}}\left(\lambda_{\text{s}},\lambda_{\text
 {bg}}|H_{\text{1}}\right)= \nonumber \\ & & \left\{\det\left[I
 \left(\lambda_{\text{s}},\lambda_{\text{bg}}|H_{\text{1}}
 \right)\right]\right\}^{\frac{1}{2}}=\sqrt{\frac{1}{\lambda_{
 \text{bg}}\left(\alpha\lambda_{\text{bg}}+\lambda_{\text{s}}
 \right)}}.
\end{eqnarray}

\section{\label{app:master1proof}Calculation of the probability of 
$H_{\text{0}}$}
For the calculation of Eqn. \ref{eqn:master1} one must solve three 
parts. First:
\begin{eqnarray}
 &  & \int_{0}^{\infty}P\left(N_{\text{on}},N_{\text{off}}|
 \lambda_{\text{bg}},H_{\text{0}}\right)  P_{\text{0}}
 \left(\lambda_{\text{bg}}|H_{\text{0}}\right)d\lambda_{\text
 {bg}}\nonumber\\ &  & =\int_{0}^{\infty}\frac{e^{-\lambda_{
 \text{bg}}\left(1+\alpha\right)}\lambda_{\text{bg}}^{N_{\text
 {off}}}\left(\alpha\lambda_{\text{bg}}\right)^{N_{\text
 {on}}}}{N_{\text{on}}!N_{\text{off}}!}\sqrt{\frac{1+\alpha}{
 \lambda_{\text{bg}}}}d\lambda_{\text{bg}}\nonumber\\ &  & =
 \frac{\alpha^{N_{\text{on}}}\left(1+\alpha\right)^{-N_{\text
 {on}}-N_{\text{off}}}\Gamma\left(\frac{1}{2}+N_{\text{on}}+N_{
 \text{on}}\right)}{N_{\text{off}}!N_{\text{on}}!},
\label{eqn:evidencea}
 \end{eqnarray}
and second:
\begin{eqnarray}
 &  & \int_{0}^{\infty}\int_{0}^{\infty}P\left(N_{\text{on}},N_{
 \text{off}}|\lambda_{\text{s}},\lambda_{\text{bg}},H_{\text{1}}
 \right)P_{\text{0}}\left(\lambda_{\text{s}},\lambda_{\text
 {bg}}|H_{\text{1}}\right)d\lambda_{\text{s}}d\lambda_{\text
 {bg}} \nonumber \\ &  & =\int_{0}^{\infty}P_{\text{P}}\left(N_{
 \text{off}}|\lambda_{\text{bg}}\right)\sqrt{\frac{1}{\lambda_{
 \text{bg}}}}  \nonumber \\ &  & \times\int_{0}^{\infty}P_{\text
 {P}}\left(N_{\text{on}}|\lambda_{\text{s}}+\alpha\lambda_{
 \text{bg}}\right)\sqrt{\frac{1}{\lambda_{\text{s}}+
 \alpha\lambda_{\text{bg}}}}d\lambda_{\text{s}}d\lambda_{\text
 {bg}} \nonumber \\ &  & =\int_{0}^{\infty}P_{\text{P}}\left(N_{
 \text{off}}|\lambda_{\text{bg}}\right)\sqrt{\frac{1}{\lambda_{
 \text{bg}}}} \frac{\Gamma\left(\frac{1}{2}+N_{\text{on}},
 \alpha\lambda_{\text{bg}}\right)}{N_{\text{on}}!}d\lambda_{
 \text{bg}}.
 \end{eqnarray}
$\Gamma\left(a,z\right)$ stands for the upper incomplete gamma function. 
The remaining integral with respect to $\lambda_{\text{bg}}$ yields:
\begin{eqnarray}
 &  & =\frac{1}{N_{\text{on}}!N_{\text{off}}!} \frac{2\alpha^{-
 \frac{1}{2}-N_{\text{off}}}}{1+2N_{\text{off}}} \Gamma\left(1+N_
 {\text{on}}+N_{\text{off}}\right)  \label{eqn:evidenceb}\\ &  & 
 \times _{2}\F_{1}\left(\frac{1}{2}+N_{\text{off}},1+N_{\text
 {on}}+N_{\text{off}};\frac{3}{2}+N_{\text{off}};-\frac{1}{\alpha}
 \right). \nonumber
\end{eqnarray}
By inserting Eqn. \ref{eqn:evidencea} and \ref{eqn:evidenceb} into 
Eqn. \ref{eqn:master1} and simplifying one finds the solution
\begin{equation}
P\left(H_{\text{0}}|N_{\text{on}},N_{\text{off}}\right) = \frac{\gamma}{\gamma+\nicefrac{c_1}{c_0}\delta} ,
\label{eqn:stepone}
\end{equation}
\begin{eqnarray}
\gamma & := & \left(1+2N_{\text{off}}\right) \alpha^{\nicefrac
{1}{2}+N_{\text{on}}+N_{\text{off}}} \\ &  & \times 
\Gamma\left(\nicefrac{1}{2}+N_{\text{on}}+N_{\text{off}}\right) 
,\nonumber\\ \delta & := & 2\left(1+\alpha\right)^{N_{\text{on}}+N_{
\text{off}}} \Gamma\left(1+N_{\text{on}}+N_{\text{off}} \right) 
\\ &  & \times _{2}\F_{1}\left(\nicefrac{1}{2}+N_{\text{off}},1+N_{
\text{on}}+N_{\text{off}};\nicefrac{3}{2}+N_{\text{off}} ;-
\nicefrac{1}{\alpha}\right) .\nonumber
\end{eqnarray}
The constant fraction $\nicefrac{c_1}{c_0}$ is calculated with Eqn. 
\ref{eqn:allenextra}, inserting the above results. Equation 
\ref{eqn:ts} follows. 
\section{\label{app:sigstr}Calculation of the marginalized 
posterior for the signal parameter}
The denominator of Eqn. \ref{eqn:marg1} is given in Eqn. \ref
{eqn:evidenceb} and does not depend on the parameters $\lambda$. The 
problem is therefore reduced to calculating the integral
\begin{eqnarray}
 &  & \int_{0}^{\infty}P\left(N_{\text{on}},N_{\text{off}}|
 \lambda_{\text{s}},\lambda_{\text{bg}},H_{\text{1}}\right)  P_{
 \text{0}}\left(\lambda_{\text{s}},\lambda_{\text{bg}}|H_{\text
 {1}}\right)d\lambda_{\text{bg}}=\nonumber\\ &  & \int_{0}^{
 \infty}P_{\text{P}}\left(N_{\text{off}}|\lambda_{\text{bg}}
 \right)  P_{\text{P}}\left(N_{\text{on}}|\lambda_{\text{s}}+
 \alpha\lambda_{\text{bg}}\right) \nonumber\\ &  & \times \sqrt
 {\frac{1}{\lambda_{\text{bg}}(\lambda_{\text{s}}+
 \alpha\lambda_{\text{bg}})}}d\lambda_{\text{bg}}=\frac{e^{-
 \lambda_{\text{s}}}}{N_{\text{on}}!N_{\text{off}}!} \nonumber\\ 
 &  & \times \int_{0}^{\infty}\left(\lambda_{\text{s}}+
 \alpha\lambda_{\text{bg}}\right)^{N_{\text{on}}-\frac{1}{2}} 
 \lambda_{\text{bg}}^{N_{\text{off}}-\frac{1}{2}}  e^{-\lambda_{
 \text{bg}}\left(1+\alpha\right)}d\lambda_{\text{bg}} ,
\end{eqnarray}
which resembles the integral representation of the Tricomi confluent 
hypergeometric function (see, for instance \citep{NIST:DLMF}, Eqn. 
(13.4.4)). By substituting the integration variable $\lambda_{\text{bg}}$
with
\begin{equation}
 \lambda=\frac{\alpha\lambda_{\text{bg}}}{\lambda_{\text{s}}} ,
\end{equation}
one finds the result
\begin{eqnarray}
 &  & \frac{e^{-\lambda_{\text{s}}}}{N_{\text{on}}!N_{\text
 {off}}!}\lambda_{\text{s}}^{N_{\text{on}}-\frac{1}{2}}
 \left(\frac{\lambda_{\text{s}}}{\alpha}\right)^{N_{\text
 {off}}+\frac{1}{2}} \Gamma\left(\frac{1}{2}+N_{\text{off}}
 \right)  \nonumber\\ &  & \times \U\left[\frac{1}{2}+N_{\text
 {off}},1+N_{\text{on}}+N_{\text{off}},\left(1+\frac{1}{\alpha}
 \right)\lambda_{\text{s}}\right] .
\label{eqn:marg3}
 \end{eqnarray}
Equations \ref{eqn:marg3} and \ref{eqn:evidenceb} put together in 
Eqn. \ref{eqn:marg1}, simplified with Eqn. \ref{eqn:poisson} and 
the regularized hypergeometric function 
$_{2}\tilde{\F}_{1}\left(a,b;c;z\right) = 
\nicefrac{_{2}\F_{1}\left(a,b;c;z\right)}{\Gamma\left(c\right)}$ give 
the final result for the marginalized posterior for 
$\lambda_{\text{s}}$:
\begin{eqnarray}
 & & P\left(\lambda_{\text{s}}|N_{\text{on}},N_{\text{off}},H_{
 \text{1}}\right) = P_{\text{P}}\left(N_{\text{on}}+N_{\text
 {off}}|\lambda_{\text{s}}\right)  \\ & & \times \frac{\U\left[
 \nicefrac{1}{2}+N_{\text{off}},1+N_{\text{off}}+N_{\text{on}},
 \left(1+\nicefrac{1}{ \alpha}\right)\lambda_{\text{s}}\right]
 }{_{2}\tilde{\F}_{1}\left(\nicefrac{1}{2}+N_{\text{off}} ,1+N_{
 \text{off}}+N_{\text{on}};\nicefrac{3}{2}+N_{\text{off}};-
 \nicefrac{1}{\alpha}\right)} . \nonumber 
\end{eqnarray}



\bibliography{OnTheOnOffProblem_aastex}



\clearpage






\end{document}